\documentclass[utf8]{modfrontiersFPHY} 

\usepackage{url,hyperref,lineno,microtype,subcaption}
\usepackage[onehalfspacing]{setspace}
\usepackage{longtable}
\usepackage{color}
\newcommand{\hl}{} 

\def\firstAuthorLast{Choi {et~al.}} 
\def\Authors{JinCheol Choi\,$^1$, Donghuan Lu\,$^2$, Mirza Faisal Beg\,$^2$, Jinko Graham\,$^1$ and Brad McNeney$^{*1,3}$ for the Alzheimer’s Disease Neuroimaging Initiative}
\def\Address{$^{1}$Department of Statistics and Actuarial Science, Simon Fraser University, Burnaby, BC, Canada \\
$^{2}$School of Engineering Science, Simon Fraser University, Burnaby, BC, Canada\\
$^{3}$Data used in preparation of this article were obtained from the 
Alzheimer’s Disease Neuroimaging Initiative (ADNI) database
(adni.loni.usc.edu). As such, the investigators within the ADNI contributed to 
the design and implementation of ADNI and/or
provided data but did not participate in analysis or writing of this report. 
A complete listing of ADNI investigators can be found
at: \url{http://adni.loni.usc.edu/wp-content/uploads/how_to_apply/ADNI_Acknowledgement_List.pdf}
}
\def\corrAuthor{Brad McNeney\\ Statistics and Actuarial Science\\ Simon Fraser University\\ Burnaby, BC\\ V5A 1S6\\ Canada \\ +1-778-782-4815}

\def\corrEmail{mcneney@sfu.ca \\ \bigskip Keywords: RV coefficient, multivariate correlation, Alzheimer's disease, shrinkage estimation, graphical display}

\begin{document}
\onecolumn
\firstpage{1}

\title[The Contribution Plot]{The Contribution Plot: Decomposition and Graphical Display of the RV Coefficient, with Application to Genetic and Brain Imaging Biomarkers of Alzheimer's Disease\\ \bigskip Running head: The Contribution Plot \\ \bigskip} 

\author[\firstAuthorLast ]{\Authors} 

\address{} 
\correspondance{} 

\extraAuth{}

\maketitle

\clearpage

\begin{abstract}

\section{Alzheimer's disease (AD) is a chronic neurodegenerative disease that causes memory loss 
and decline in cognitive abilities. AD is the sixth leading cause of death in the 
United States, affecting an estimated 5 million Americans. 
To assess the association between multiple genetic variants and 
multiple measurements of structural changes in the brain a recent study of
AD used a multivariate measure of linear dependence, the 
RV coefficient. 
The authors 
decomposed the RV coefficient into contributions from individual variants and 
displayed these contributions graphically. We investigate the properties of 
such a ``contribution plot'' in terms of an underlying linear model, and 
discuss estimation of the components of the plot when the correlation signal 
may be sparse. The contribution plot is applied to simulated data 
and to genomic and brain imaging data from the Alzheimer's Disease Neuroimaging Initiative. \\
}

\end{abstract}

\section{Introduction}

Alzheimer's disease (AD) is a neurodegenerative disorder. As a type of dementia, 
it is a neurological dysfunction that is irreversible, neurodegenerative and
progressive, causing memory loss and the decline of cognitive function. AD usually 
occurs in older people and 
is considered to be a complex disease driven by a combination of genetic and 
environmental factors.
More than 5 million Americans 
suffer from AD and it is ranked as \hl{the sixth largest cause of mortality} in the 
USA \citep{alzheimer20172017}.

The Alzheimer's Disease Neuroimaging Initiative (ADNI) is a longitudinal, 
multi-site study that started in 2004 to understand the onset, progression, 
and etiology of AD. One of the ADNI objectives is to 
identify associations between genetic and brain-imaging biomarkers of 
AD \citep{weiner2013alzheimer}.
Neuroimaging studies such as ADNI feature multivariate datasets, typically 
comprised of large numbers of genotypes and phenotypes. 
For example, the dataset in \cite{szefer2017multivariate} consisted of 
75,181 single nucleotide polymporphism (SNP) genotypes and 56 
brain phenotypes derived from MRI scans.

To measure the
association between multivariate datasets, many 
different correlation coefficients have been introduced.
One of the most popular 
is the RV coefficient, which measures the linear association between two datasets
by estimating the population
vector correlation coefficient $\rho_V$ \citep{JosseHolmes16}.
When both datasets consist of a single variable, RV is 
the squared Pearson-correlation coefficient and $\rho_V$ is the
squared population-correlation coefficient.

\begin{figure}[h]
  \centering
  \includegraphics[width=6in]{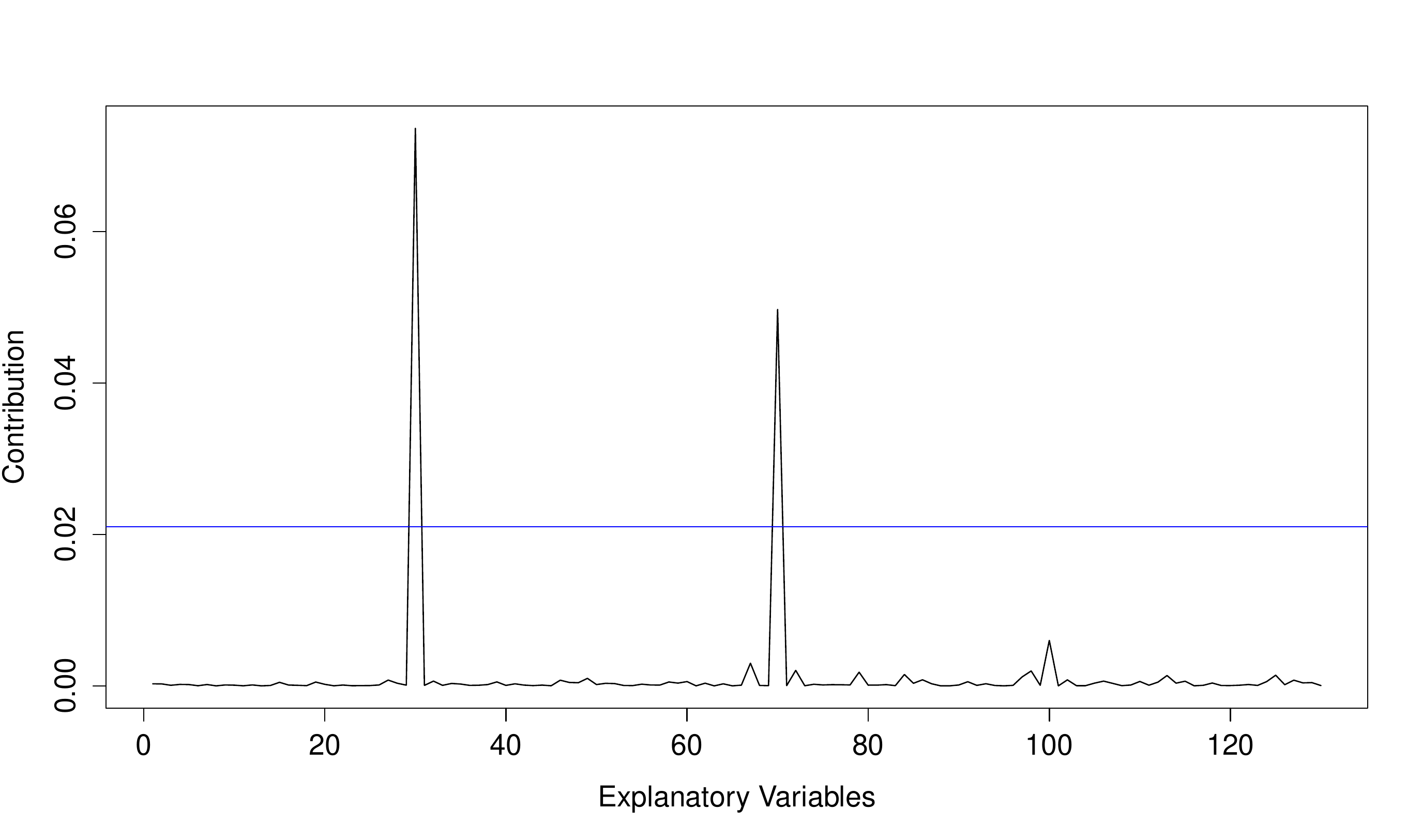}
  \caption
   {Example contribution plot for data simulated as described in Section \ref{sec:sim} (example dataset 3), using the methods of Section~\ref{sec:contribplot}. The vertical axis is the contribution of each explanatory variable to a modified RV coefficient designed to identify sparse correlation signals (Section~\ref{sec:contribplot}). The horizontal axis is the index of the explanatory variables. The horizontal line is the estimated 95th percentile of the distribution 
of the maximum contributions under no association, where the maximum is over all explanatory variables.
The estimate is based on an empirical null 
distribution from 5,000 data sets in which the rows of the matrix of
explanatory variables are permuted. Individual contributions 
that exceed the 95\% threshold are considered noteworthy.}
  \label{fig:excontrib}
\end{figure}

In \cite{szefer2017multivariate}, the RV coefficient is used to summarize 
the multivariate association between brain phenotypes and SNPs in AD linkage regions.
These authors performed a test of the null hypothesis $\rho_{V} = 0$ {\it vs} the 
alternative hypothesis $\rho_{V} > 0$ and rejected the null hypothesis. 
In a {\it post hoc} investigation, they decomposed the RV coefficient into 
contributions from each SNP and plotted the result \citep[][Figure 5]{szefer2017multivariate}.
An example contribution plot using the methods described in Section~\ref{sec:contribplot} 
of this article is given in Figure~\ref{fig:excontrib}.
The plot suggests that the association between the multivariate data matrices
of explanatory and response variables is driven by the 30th and 70th explanatory 
variables.

In this report we investigate the properties of 
the contribution plot in terms of an underlying linear model, and 
discuss estimation of the components of the plot when the correlation signal 
may be sparse. The contribution plot is applied to 
simulated datasets and to 
genetic and brain-imaging data from the ADNI study.

\section{Material and Methods}

\subsection{The Contribution plot}
\label{sec:contribplot}

In this section, we define the RV coefficient and its population counterpart, 
the multivariate correlation coefficient $\rho_V$, following 
\cite{JosseHolmes16}. 
Our intended use of the RV coefficient is to investigate correlations between
matrices of genetic marker genotypes and brain phenotypes, and our descriptions
will be in those terms, though the methods apply in any multivariate setting.
We decompose $\rho_V$ 
into contributions from each genetic marker, and study the form of such contributions 
under a multivariate linear model for brain phenotypes given genomic data. 
Finally, we discuss shrinkage estimation of the contributions that may be useful when the correlation 
signal is sparse. By sparse we mean few non-zero pairwise correlations between 
genotypes and phenotypes.  

Let $X=(X_1,\ldots,X_p)$ denote a random vector of $p$ explanatory variables and
$Y=(Y_1,\ldots,Y_q)$ denote a random vector of $q$ response variables. 
A measure of population correlation between $X$ and $Y$ is \citep{Escoufier73}
\begin{equation}
\rho_V(X,Y) = \frac{\displaystyle\sum_{k=1}^p \sum_{l=1}^q\mathrm{Cov}^2(X_k,Y_l)}{\displaystyle\sqrt{\sum_{k=1}^p \sum_{l=1}^p\mathrm{Cov}^2(X_k,X_l) \sum_{k=1}^q \sum_{l=1}^q\mathrm{Cov}^2(Y_k,Y_l)}},
\label{eqn:rhoV}
\end{equation}
where Cov() denotes population covariance. The coefficient $\rho_V$ may 
be viewed as an extension of the squared population correlation to the 
multivariate setting. 

Suppose we have $n$ independent and identically distributed realizations of 
$X$ and $Y$, arranged row-wise as data matrices 
$\textbf{X}$($n\times p$) and $\textbf{Y}$($n\times q$), respectively. 
Let $X_{\cdot k}$  denote the $k$th column of $\textbf{X}$; i.e., the 
vector of genotypes for genetic marker $k$. Similarly, let $Y_{\cdot l}$ 
denote the $l$th column of $\textbf{Y}$; i.e., the vector of
measurements for phenotype $l$. The multivariate correlation coefficient
in equation~(\ref{eqn:rhoV}) can be estimated by the RV coefficient, 
obtained by replacing population covariances such as $\mathrm{Cov}(X_k,Y_l)$ 
with their sample counterparts $\mathrm{cov}(X_{\cdot k},Y_{\cdot l})$:
\begin{equation} \label{cov_eq}
RV(\textbf{X},\textbf{Y})=\displaystyle \frac{\displaystyle\sum_{k=1}^{p}\displaystyle\sum_{l=1}^{q}\mathrm{cov}^{2}(X_{.k},Y_{.l})}{\sqrt{\displaystyle\sum_{k=1}^{p}\displaystyle\sum_{l=1}^{p}\mathrm{cov}^{2}(X_{.k},X_{.l})\displaystyle\sum_{k=1}^{q}\displaystyle\sum_{l=1}^{q}\mathrm{cov}^{2}(Y_{.k},Y_{.l})}}.
\end{equation}
Reference \cite{JosseHolmes16} and Appendix A of \cite{Choi2018} give 
alternate forms of the RV coefficient.

From equation \eqref{cov_eq}, the contribution of the $k$th genetic marker to the 
RV coefficient is proportional to
\begin{equation} \label{eq:estContrib}
\hat{\mathcal{C}}_k = \sum_{l=1}^q \mathrm{cov}^2(X_{.k},Y_{.l}).
\end{equation}
The notation $\hat{\mathcal{C}}_k$ reflects the fact that the contribution of 
genetic marker $k$ to the RV coefficient is an estimate of a corresponding contribution 
to $\rho_V(\textbf{X},\textbf{Y})$:
\begin{equation}
\mathcal{C}_{k}= \sum_{l=1}^q\mathrm{Cov}^2(X_k,Y_l).
\label{contri_eq}
\end{equation}
The covariances that comprise $\mathcal{C}_k$ can be derived under a 
linear model for the association between $X$ and $Y$. 
Such a model is consistent with the RV coefficient measuring 
the linear relationship between two multidimensional datasets. In fact,
\begin{equation} \label{contri_eq_3}
    \mathcal{C}_k
    =\displaystyle\sum_{l=1}^{q}\left\{ \beta_{kl} \text{Var}(X_k) + \displaystyle\sum_{k'\not= k}\beta_{k'l}\text{Cov}(X_k,X_{k'}) \right\}^{2},
\end{equation}
where $\beta_{kl}$ is the coefficient of $X_k$ in the regression of $Y_l$ on $X$ \cite{Choi2018}.
Equation~\eqref{contri_eq_3} shows that 
$\mathcal{C}_k$ depends on not only the regression coefficients, but 
also the variance of $X_k$ and the covariances between $X_k$ and the other 
components of $X$.  Some simplification of the contributions is obtained 
by scaling each $X_k$ by its standard deviation, so that the variance 
terms become one and covariances become correlations. 
Letting $X^*$ and $Y^*$ denote the standardized variables, 
the contribution of genetic marker $k$ to $\rho_V(X^*,Y^*)$ is 
\begin{equation} \label{contri_equ}
    \mathcal{C}_k^{*}  =\displaystyle\sum_{l=1}^{q}\left\{ \beta^*_{kl} + \displaystyle\sum_{k'\not= k}\beta^*_{k'l}\text{Cor}(X^*_k,X^*_{k'}) \right\}^{2},
\end{equation}
where $\beta^*_{kl}$ is the coefficient of the standardized $X^*_k$ in the regression 
of the standardized $Y^*_l$ on $X^*$.
Thus, genetic marker $k$ makes a non-zero 
contribution to $\rho_V(X^*,Y^*)$ if it is directly associated with 
one or more $Y_l$ (i.e., $\beta_{kl} \not= 0$ for one or more $l$) {\em or} 
if it is correlated with one or more $X_{k'}$ that is/are directly 
associated with one or more $Y_l$ (i.e., there is a $k'$ such 
that Cor$(X_k,X_{k'}) \not= 0$ and an $l$ such that $\beta_{k'l} \not= 0$). 
Interestingly, a genetic marker's indirect associations with phenotypes
do not play a role in determining its contribution; we return to this 
point in the analyses of simulated data.

We now turn to estimation of the contributions to the RV coefficient. 
The contribution from the $k$th genetic marker is
\begin{equation} \label{contri_equ_cor_form}
\hat{\mathcal{C}}_k^* =  \displaystyle\sum_{l=1}^{q}\mathrm{cor}^{2}(X^*_{.k},Y^*_{.l}),
\end{equation}
a sum of squared sample correlations. Our studies of simulated data 
(Section \ref{sec:simres}) suggest that when the correlation signal is sparse, in the sense 
that there are few truly non-zero correlations, and the sample 
size is modest compared to the number of phenotypes, sampling 
error in estimates of truly {\em zero} correlations can obscure 
the signal of the truly non-zero correlations. A solution is to 
raise the squared correlations to a power, $\alpha$; i.e., we consider the contributions
\begin{equation} \label{contri_stnd_alpha_eq}
\hat{\mathcal{C}}_k^*(\alpha) =  \displaystyle\sum_{l=1}^{q}\mathrm{cor}^{2\alpha}(X^*_{.k},Y^*_{.l})
\end{equation}
to a modified RV coefficient
\begin{equation}
RV(\textbf{X}^*,\textbf{Y}^*|\alpha) \propto \displaystyle\sum_{k=1}^p\displaystyle\sum_{l=1}^{q}\mathrm{cor}^{2\alpha}(X^*_{.k},Y^*_{.l})
\label{eqn:modRV}
\end{equation}
for $\alpha \geq 1$.  Raising correlations to powers larger than 2 has 
the effect of differentially shrinking all estimates toward zero, with 
estimates near zero shrunken more than those near one. 
Independently, \cite{xu2017adaptive} arrived at the same modified RV 
coefficient in the context of testing the null hypothesis 
$H_0:\rho_V(X^*,Y^*)=0$ {\it versus} the alternative hypothesis 
$H_1:\rho_V(X^*,Y^*) > 0$. In their sum-of-powered-correlations
test, SPC($\alpha$), they employ $RV(\textbf{X}^*,\textbf{Y}^*|\alpha)$ 
as a test statistic and assess its significance with a 
Monte Carlo permutation test. They also suggest 
an adaptive test (aSPC), in which the test statistic is a minimum p-value 
for the SPC($\alpha$) test over a grid of powers. 
Though testing is not the focus of this project, we make use of their 
minimum-p-value idea to select a power $\alpha$ for the contribution 
plot. In particular, our 
contribution plot is of contributions $\hat{\mathcal{C}}_k^*(\alpha_m)$ 
for the power \hl{$\alpha_m$ that minimizes the p-value of the test based 
on $RV(\textbf{X}^*,\textbf{Y}^*|\alpha)$, for values of $\alpha$ 
on a grid}. In our study we chose the grid $\alpha=1,2,3$ or 4.
R \cite{Rcore} code to implement the contribution plot is given in the Appendix.

\subsection{Simulated Data Settings}
\label{sec:sim}

We applied the contribution plot to 
simulated multivariate datasets consisting of a matrix of 
explanatory variables $\textbf{X}$ and a matrix of response variables $\textbf{Y}$.
Here we summarize results from three datasets simulated to represent
no or sparse association. 
To investigate the effect of correlation
among explanatory variables and correlation among response variables
on the properties of the contribution plot, we simulated data
with and without these correlations, as described next.

Simulated datasets consisted of 
$p=130$ explanatory variables and 
$q=25$ response variables on $n=100$ subjects.
We simulated from a multivariate multiple-regression model 
$$
\textbf{Y}= \textbf{XB}+\textbf{E},
$$
in which
$\textbf{Y}_{n \times q}$ is a matrix of response variables,
$\textbf{X}_{n \times p}$ is a matrix of explanatory 
variables generated from $MVN(0,\Sigma^X)$,
$\textbf{B}_{p \times q}$ is a coefficient matrix and 
$\textbf{E}_{n \times q}$ is an error matrix generated from $MVN(0,\Sigma^E)$.

In our simulation model, we vary the parameters
$\Sigma^X$, $\Sigma^E$, 
and $\textbf{B}$. Let $I_p$ and $I_q$ denote the $p\times p$ and 
$q\times q$ identity matrices. 
We summarize results from three datasets simulated under the following
parameter values:
\begin{itemize}
\item {\bf Dataset 1:} No assocaitions
\begin{itemize}
\item $\Sigma^X = I_p$, 
\item $\Sigma^E=I_q$, 
\item $B_{ij}=0$ for all $i$ and $j$. 
\end{itemize}
\item {\bf Dataset 2:} Sparse association; correlated explanatory variables $X_{25},\ldots,X_{35}$
\begin{itemize}
\item $\Sigma^X_{i,j}$=0.9 for $25 \leq i,\; j \leq 35$ and $i\not= j$, with
all diagonal entries equal to 1 and all other entries equal to 0.
\item $\Sigma^E = I_q$
\item $B_{30,1}=B_{70,10}=1$ so that $X_{30}$ and $X_{70}$ are causally associated
with $Y_1$ and $Y_{10}$, respectively.
\end{itemize}
\item {\bf Dataset 3:} Sparse associations; correlated errors $E_1,\ldots,E_{15}$, and hence correlated responses
\begin{itemize}
  \item $\Sigma^X = I_p$
  \item $\Sigma^E_{i,j}$=0.9 for $1 \leq i,\; j \leq 15$ and $i\not= j$, with
  all diagonal entries of $\Sigma^E$ equal to 1 and all other entries equal to 0.
  \item $B_{30,1}=B_{70,10}=1$ so that $X_{30}$ and $X_{70}$ are causally associated
  with $Y_1$ and $Y_{10}$, respectively.
\end{itemize}
\end{itemize}

Further simulation settings were considered in \cite{Choi2018}, Chapter 4, but
we do not present them here.

\subsection{ADNI Data Description}

In this section we describe the ADNI data used to illustrate
the contribution plot. 

\subsubsection{Subjects}
Both SNP and brain image data considered in this analysis were from 
the ADNI Phase 1 (ADNI-1) study that was run in the years 2004 through 2009. 
Our interest was in genetic variation that predicts structural differences in the 
brain \emph{before} subjects experience memory loss. 
Hence we considered data from the 200 cognitively normal (CN) subjects only.
Further details about the ADNI-1 
study design are available on the ADNI website \url{http://adni.loni.usc.edu/study-design/}.

\bigskip

\begin{table}[htb]
\centering
\begin{tabular}{ccc|ccc}
\hline
Chromosome & Gene   & No. & Chromosome & Gene    & No. \\ \hline
1          & CHRNB2 & 1   & 10         & SORCS1  & 94  \\
1          & CR1    & 15  & 10         & TFAM    & 6   \\
1          & ECE1   & 39  & 11         & GAB2    & 19  \\
1          & MTHFR  & 10  & 11         & PICALM  & 23  \\
1          & TF     & 3   & 11         & SORL1   & 33  \\
2          & BIN1   & 12  & 15         & ADAM10  & 19  \\
2          & IL1A   & 2   & 17         & ACE     & 7   \\
2          & IL1B   & 1   & 17         & GRN     & 1   \\
6          & NEDD9  & 69  & 17         & THRA    & 3   \\
6          & PGBD1  & 6   & 17         & TNK1    & 3   \\
6          & TNF    & 1   & 19         & APOE    & 1   \\
8          & CLU    & 2   & 19         & EXOC3L2 & 2   \\
9          & DAPK1  & 82  & 19         & GAPDHS  & 3   \\
9          & IL33   & 14  & 19         & LDLR    & 9   \\
10         & CALHM1 & 3   & 20         & CST3    & 1   \\
10         & CH25H  & 1   & 20         & PRNP    & 4   \\
10         & ENTPD7 & 4   & Total      &         & 493 \\ \hline
\end{tabular}
\caption{Summary of the number of SNPs in analyzed genes.}
\label{table:number_of_SNPs}
\end{table}

\subsubsection{Genotype Data}
Genotyping was performed as described in \cite{saykin2010alzheimer}. Genotypes
were processed according to standard 
quality control and imputation procedures to fill in missing 
values as described in \cite{szefer2014joint}. 
SNPs were chosen from the top 40 AD candidate genes listed on the
AlzGene database as of June 10, 2010.
After data processing, 
179 subjects with data on 493 SNPs in 33 genes remained for analysis. 
Table \ref{table:number_of_SNPs} gives a summary of gene names and the 
numbers of SNPs from each gene. SNP names are given in Appendix C of \cite{Choi2018}.

\subsubsection{Imaging Phenotype Data}
The phenotypes, as defined in \cite{wang2011identifying},
were derived from baseline MRI scans taken 
for the ADNI-1 study. The MRI measurements were of volumes 
or cortical thicknesses of 56 brain regions (Table \ref{FreeSurfer}), 
adjusted for covariates such as age, gender, education level, 
handedness and baseline intracranial volume.

\begin{footnotesize}
\begin{longtable}{llp{9cm}}
\hline
\multicolumn{1}{c}{\textbf{Phenotype ID}} & \multicolumn{1}{c}{\textbf{Measurement}} & \multicolumn{1}{c}{\textbf{Cerebral region}} \\ \hline
AmygVol & Volume & Amygdala  \\
CerebCtx & Volume & Cerebral cortex  \\
CerebWM  & Volume & Cerebral white matter \\
HippVol  & Volume & Hippocampus \\
InfLatVent & Volume & Inferior lateral ventricle  \\
LatVent  & Volume  & Lateral ventricle  \\
EntCtx   & Thickness  & Entorhinal cortex \\
Fusiform  & Thickness & Fusiform gyrus \\
InfParietal & Thickness  & Inferior parietal gyrus \\
InfTemporal & Thickness & Inferior temporal gyrus  \\
MidTemporal & Thickness  & Niddle temporal gyrus  \\
Parahipp  & Thickness   & Parahippocampal gyrus \\
PostCing  & Thickness   & Posterior cingulate  \\
Postcentral  & Thickness & Postcentral gyrus   \\
Precentral & Thickness  & Precentral gyurs  \\
Precuneus   & Thickness   & Precuneus  \\
SupFrontal  & Thickness & Superior frontal gyrus \\
SupParietal  & Thickness  & Superior parietal gyurs  \\
SupTemporal  & Thickness  & Superior temporal gyrus \\
Supramarg  & Thickness  & Supramarginal gyrus \\
TemporalPole & Thickness & Temporal pole   \\
MeanCing   & Mean thickness & Caudal anterior cingulate, isthmus cingulate, posterior cingulate, and rostral anterior cingulate  \\
MeanFront & Mean thickness & Caudal midfrontal, rostral midfrontal, superior frontal, lateral orbitofrontal, and medial orbitofrontal gyri and frontal pole \\
MeanLatTemp  & Mean thickness  & Inferior temporal, middle temporal, and superior temporal gyri \\
MeanMedTemp  & Mean thickness  & Fusiform, parahippocampal, and lingual gyri, temporal pole and transverse temporal pole \\
MeanPar & Mean thickness & Inferior and superior parietal gyri, supramarginal gyrus, and precuneus \\
MeanSensMotor & Mean thickness  & Precentral and postcentral gyri \\
MeanTemp  & Mean thickness & Inferior temporal, middle temporal, superior temporal, fusiform, parahippocampal, and lingual gyri, temporal pole and transverse temporal pole \\ \hline
\caption{Phenotype IDs and descriptions of 28 brain regions from a hemisphere, from
Table 2.1 of \cite{szefer2014joint}. Baseline structural MRI measurements of a total of 56 (= 28 $\times$
2) regions from left and right hemispheres were estimated.}
\label{FreeSurfer}
\end{longtable}
\end{footnotesize}

\subsubsection{Adjustment for Potential Confounders}
Following \cite{szefer2017multivariate}, 
phenotypes and genotypes were adjusted for ethnicity and APOE
genotypes.
Ethnicity was represented by the top 10 principal components of a 
genome-wide set of approximately independent genetic markers.
Adjusted variables were taken to be the residuals from
a linear regression on these principal components
and APOE genotype categories.

\subsubsection{Standardization}

\hl{Data cleaning and adjustment for confounders lead to a
$179\times 493$ matrix of explanatory variables $\textbf{X}$ and a
$179\times 52$ matrix of response variables $\textbf{Y}$.
Each column of $\textbf{X}$ and of $\textbf{Y}$ is 
a residual and therefore has sample mean zero.
The final step of data preparation was to standardize each 
column of $\textbf{X}$ and $\textbf{Y}$ by dividing by their standard deviation.}

\section{Results}

\subsection{Simulated Data Results}
\label{sec:simres}

We applied the contribution plot to each of the datasets simulated
as described in Section \ref{sec:sim}.
For each dataset we report the p-value for the aSPC test and
show the contribution plot. Recall that the
contribution plot is of the contributions to $RV(\textbf{X}^*,\textbf{Y}^*|\alpha_m)$
for the value $\alpha_m$ that minimizes the p-value of the SPC($\alpha$) test
over the grid of values $\alpha=1,2,3,4$.
For comparison, we also plot the contributions to 
$RV(\textbf{X}^*,\textbf{Y}^*|1)$.

\textbf{Dataset 1:} No association

\begin{figure}[ht]
  \centering
  \includegraphics[width=6in]{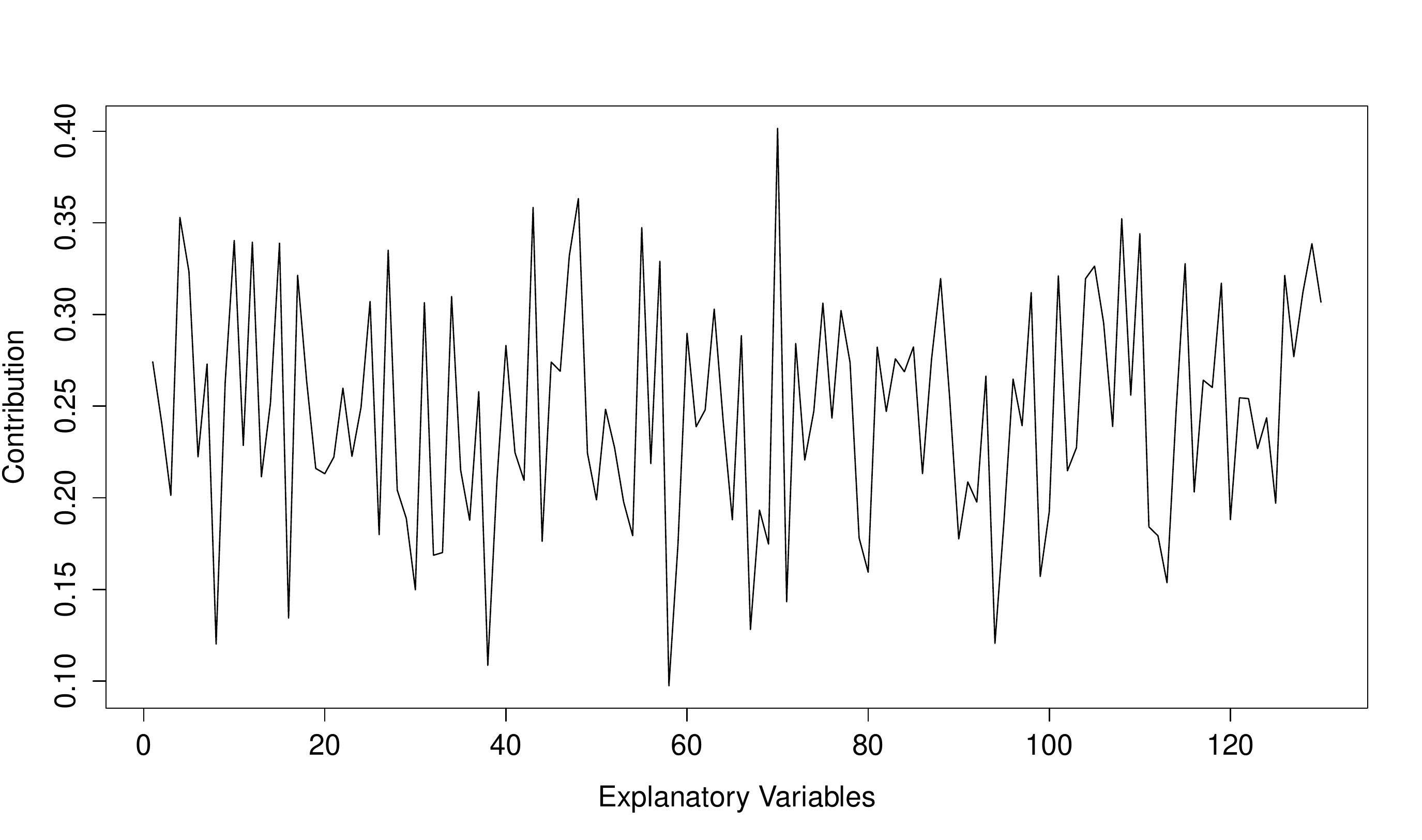}
    
  \vspace{20pt}
  
  \includegraphics[width=6in]{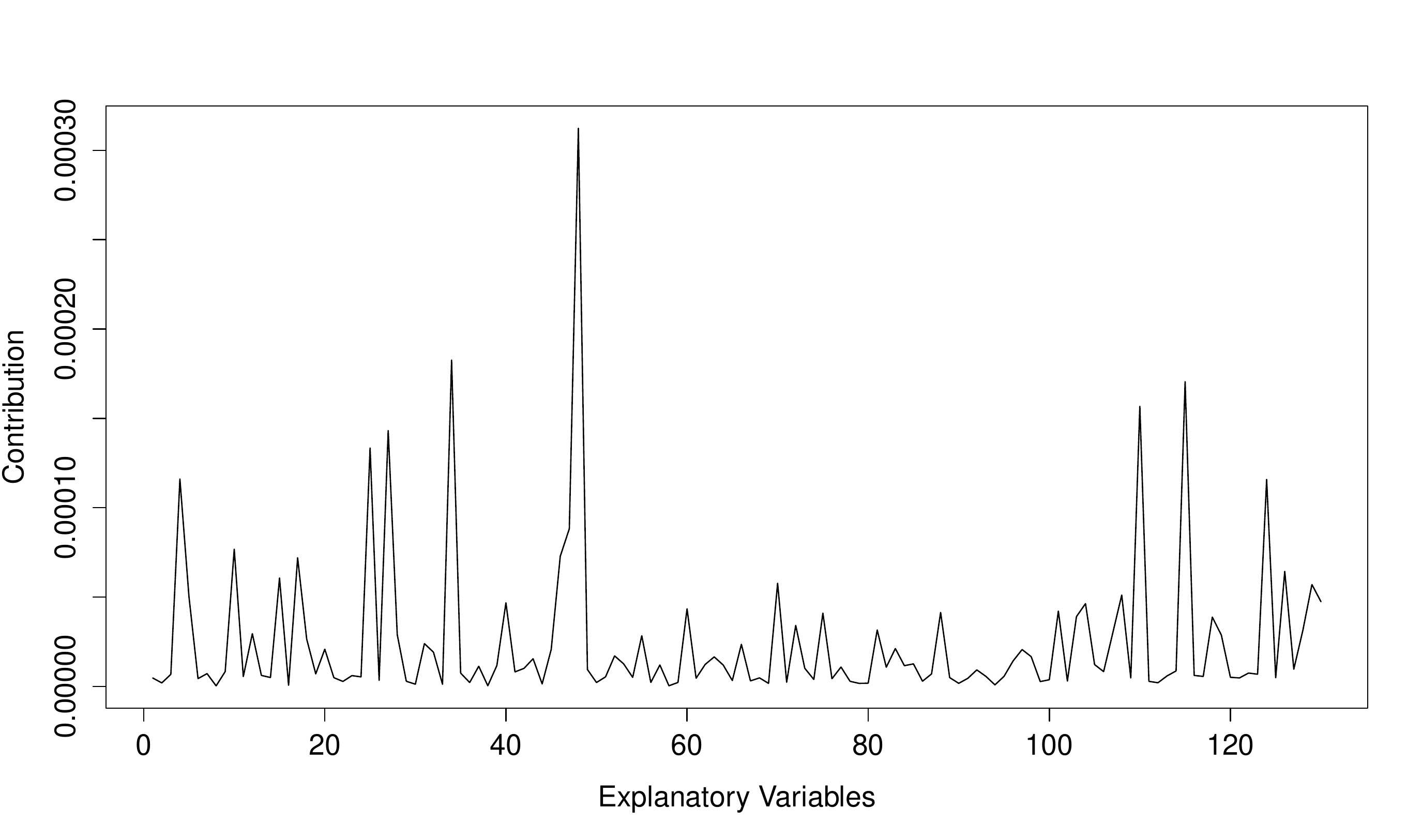}
  \caption
   {Simulation results for dataset 1 (null hypothesis). The top panel shows contributions
to $RV(\textbf{X}^*,\textbf{Y}^*|1)$ and the bottom panel shows the contribution plot (contributions to $RV(\textbf{X}^*,\textbf{Y}^*|\alpha_m)$ with $\alpha_m=4$).}
  \label{fig:sim_0}
\end{figure}

The p-value for the aSPC test on this simulated dataset is 0.5055, 
correctly suggesting no association.
Figure \ref{fig:sim_0} displays the $RV(\textbf{X}^*,\textbf{Y}^*|\alpha)$
contributions 
for $\alpha=1$ (top panel) and $\alpha_m=4$ (bottom panel). The 
significance threshold for the top panel is 0.5498 and the 
threshold for the bottom panel is 0.0059; both are outside 
the range of the vertical axes on the plots. In both panels 
there are no contributions that meet or exceed the significance 
thresholds. Thus, all contributions are considered true-negatives.

\noindent
\textbf{Dataset 2:} Sparse association; correlated explanatory variables

These data were simulated with equi-correlated explanatory variables
$X_{25},\ldots,X_{35}$. The sample correlations between these
explanatory variables ranged between 0.77 and 0.97 (median 0.91). 

\begin{figure}[h]
  \centering
  \includegraphics[width=6in]{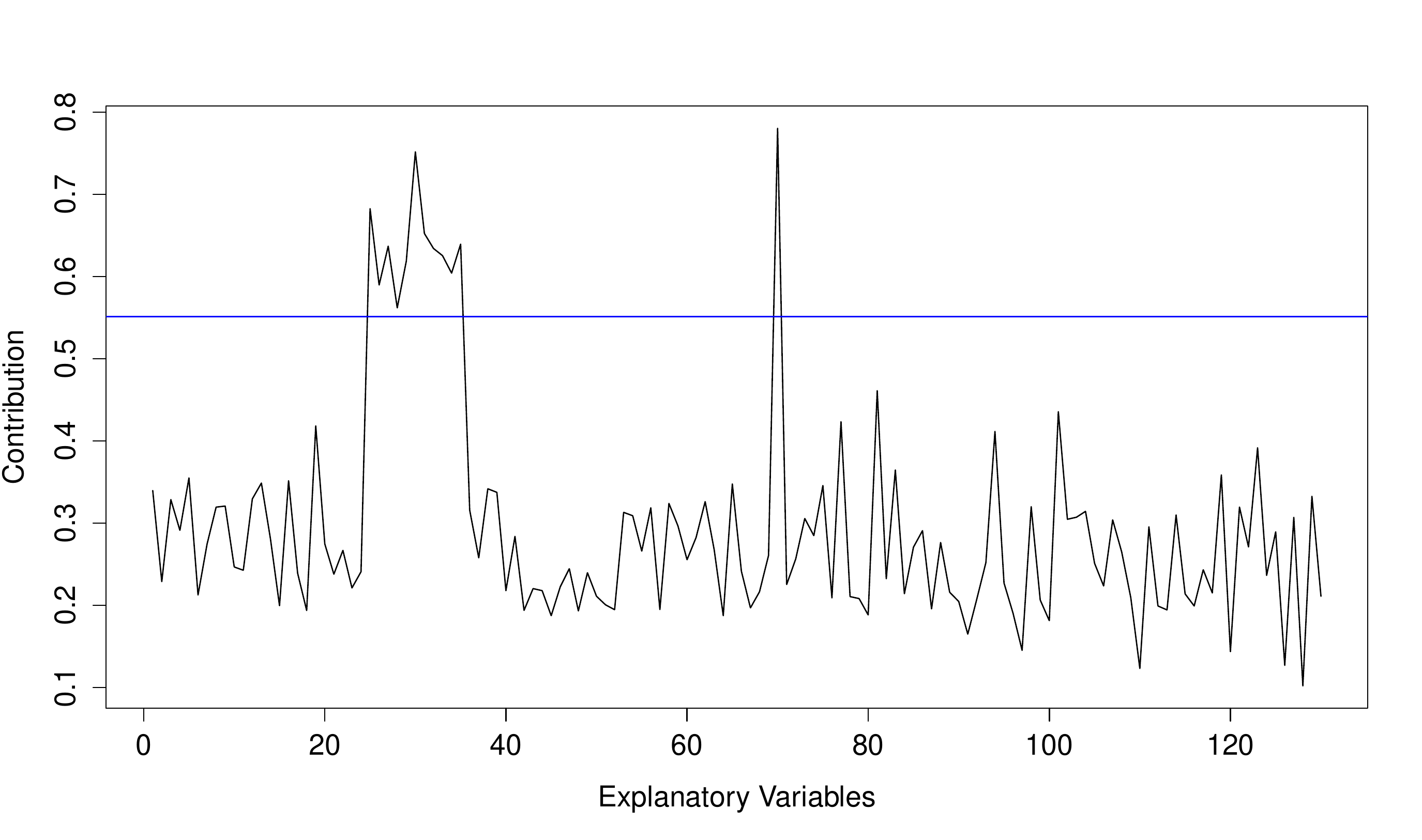}
    
  \vspace{20pt}
  
  \includegraphics[width=6in]{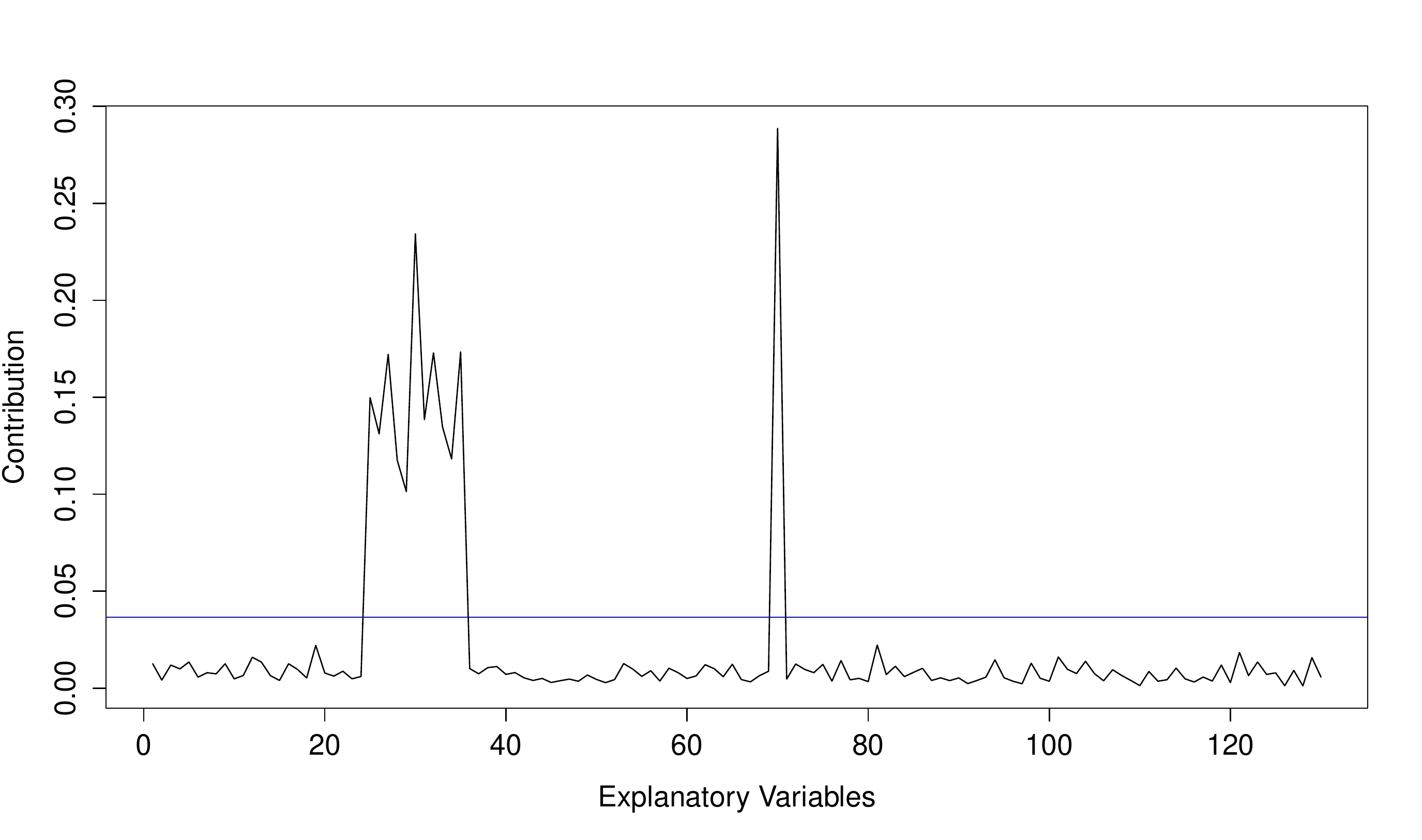}
  \caption
   {Simulation results for dataset 2 (correlated explanatory variables). The top panel shows contributions
to $RV(\textbf{X}^*,\textbf{Y}^*|1)$ and the bottom panel shows the contribution plot (contributions to $RV(\textbf{X}^*,\textbf{Y}^*|\alpha_m)$ with $\alpha_m=2$). The horizontal line indicates the 95th percentile of the maximum contributions under the permutation null distribution.}
  \label{fig:sim_1}
\end{figure}

The p-value for the aSPC test on this simulated dataset is 0.0006, 
reflecting the true association between the 30th explanatory 
variable, $X_{30}$, and the first response variable, $Y_1$, and 
between the 70th explanatory variable, $X_{70}$, and the 10th 
response variable, $Y_{10}$.
The $RV(\textbf{X}^*,\textbf{Y}^*|\alpha)$ contributions 
for $\alpha=1$ and $\alpha_m=2$ are shown in Figure \ref{fig:sim_1}. 
The broad peak of signal toward the left end of the horizontal 
axes of the plots reflects the truly-associated $X_{30}$. In 
addition to a signal at $X_{30}$, other explanatory variables 
that are correlated with $X_{30}$ have comparably-sized 
contributions, as predicted by equation \eqref{contri_equ}. 
In particular, combining the data-generating model with 
the equation for the contributions (equation \ref{contri_equ}) we obtain: 
\begin{center}
    $\mathcal{C}_i^{*} = \displaystyle\sum_{l=1}^{25}\left\{ \beta^*_{il} + \displaystyle\sum_{k'\not= i}\beta^*_{k'l}\text{Cor}(X^*_i,X^*_{k'}) \right\}^{2}$.
\end{center}
From the equation above,
one can argue that $\mathcal{C}^*_{30}=1$ and $\mathcal{C}^*_i=.81$ 
for $i=25,\ldots,29,31,\ldots,35$.
Thus we expect the observed peak of contributions at the 30th variable, surrounded by sub-peaks of 
about 80\%-peak-height from indices 25 to 35.
The narrow peak near the middle of the horizontal axes in Figure \ref{fig:sim_1} 
reflects the truly-associated $X_{70}$, which is not correlated 
with any of the other explanatory variables. 

\vspace{10pt}

\noindent
\textbf{Dataset 3:} Sparse association; correlated response variables

\vspace{10pt}

For this dataset, the response variables $Y_1,\ldots,Y_{15}$ are constructed from
equi-correlated errors $E_1,\ldots,E_{15}$. Response variable $Y_1$ is linearly related  
to explanatory variable $X_{30}$, but $Y_2,\ldots,Y_{15}$ are not related 
to any of the explanatory variables.
The linear trend in $Y_1$ reduces its correlation with $Y_2,\ldots,Y_{15}$:
sample correlations between $Y_1$ and responses $Y_2,\ldots,Y_{15}$ ranged from 0.50 to
0.71 (median 0.68), while sample correlations among $Y_2,\ldots,Y_{15}$ ranged
from 0.71 to 0.95 (median 0.92).

\begin{figure}[!htb]
  \centering
  \includegraphics[width=6in]{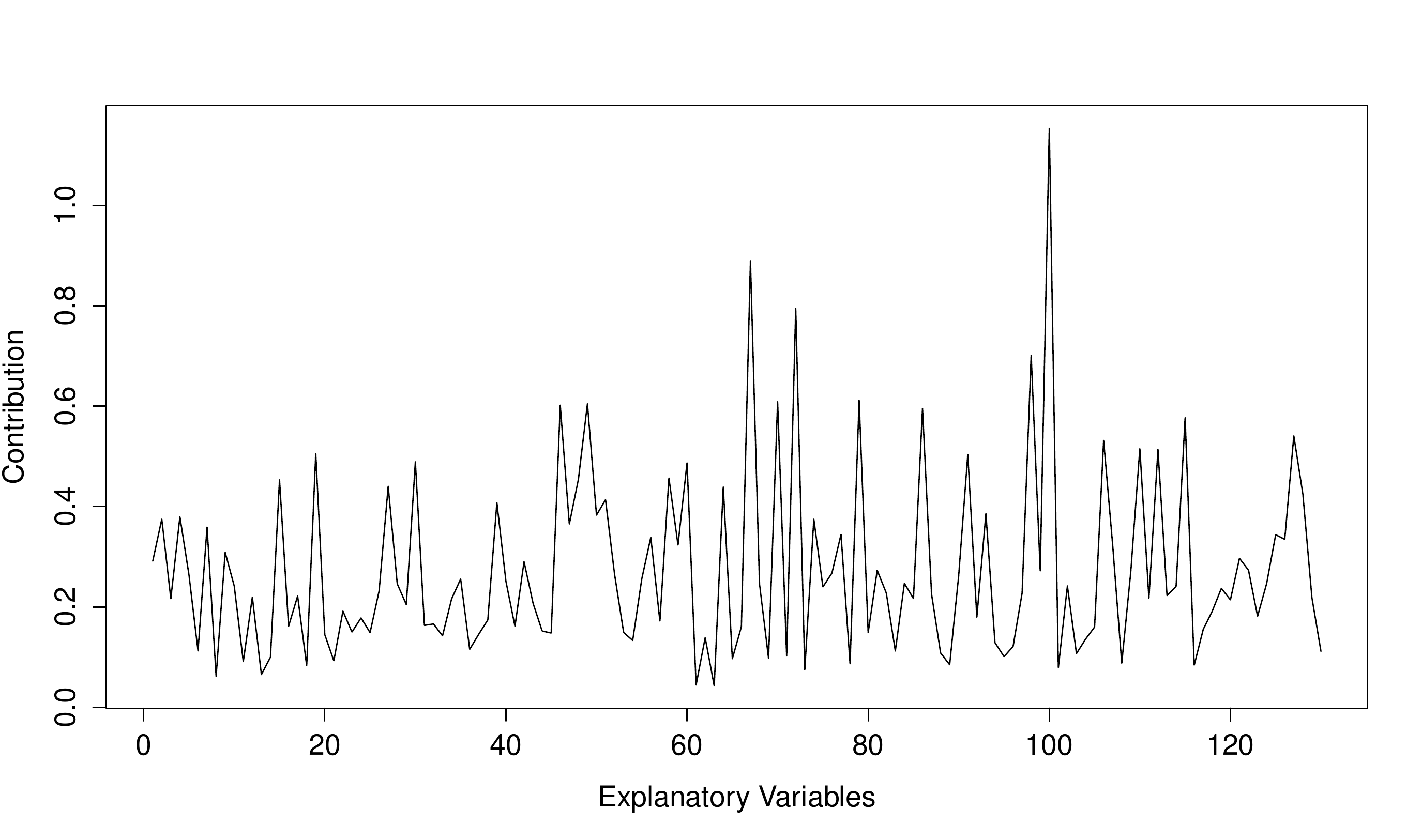}
  
  \vspace{10pt}
  
  \includegraphics[width=6in]{sm2_stnd_x_y_alpha_3}
  \caption
   {Simulation results for dataset 3 (correlated response variables). The top panel shows contributions
to $RV(\textbf{X}^*,\textbf{Y}^*|1)$ and the bottom panel shows the contribution plot (contributions to $RV(\textbf{X}^*,\textbf{Y}^*|\alpha_m)$ with $\alpha_m=3$). The horizontal line in the lower panel indicates the 95th percentile of the maximum contributions under the permutation null distribution.}
  \label{fig:sim_2}
\end{figure}

The p-value for the aSPC test on this simulated dataset is 0.0008, 
reflecting the true association between $X_{30}$ and $Y_1$ and 
between $X_{70}$ and $Y_{10}$. 
The $RV(\textbf{X}^*,\textbf{Y}^*|\alpha)$ contributions 
for $\alpha=1$ and $\alpha_m=3$
are shown in Figure \ref{fig:sim_2}. For 
contributions to $RV(\textbf{X}^*,\textbf{Y}^*|1)$ the significance 
threshold is 1.7055. In the top panel we see that none of the 
contributions exceed this threshold. The 
increased threshold in dataset 3 compared to dataset 2 is a 
consequence of the increased variance in the contributions 
$\hat{\mathcal{C}}_k^*(\alpha)= \sum_{l=1}^{q}\mathrm{cor}^{2\alpha}(X^*_{.k},Y^*_{.l})$ 
resulting from positive dependence between response variables.  
In the top panel of Figure \ref{fig:sim_2}, the peak signal is at $X_{100}$, which is not 
truly associated with any of the response variables. By contrast, 
in the contribution plot of the bottom panel,
the contributions of the two truly-associated 
variables do exceed the threshold.

\vspace{10pt}

\begin{figure}[!htb]
  \centering
  \includegraphics[width=6in]{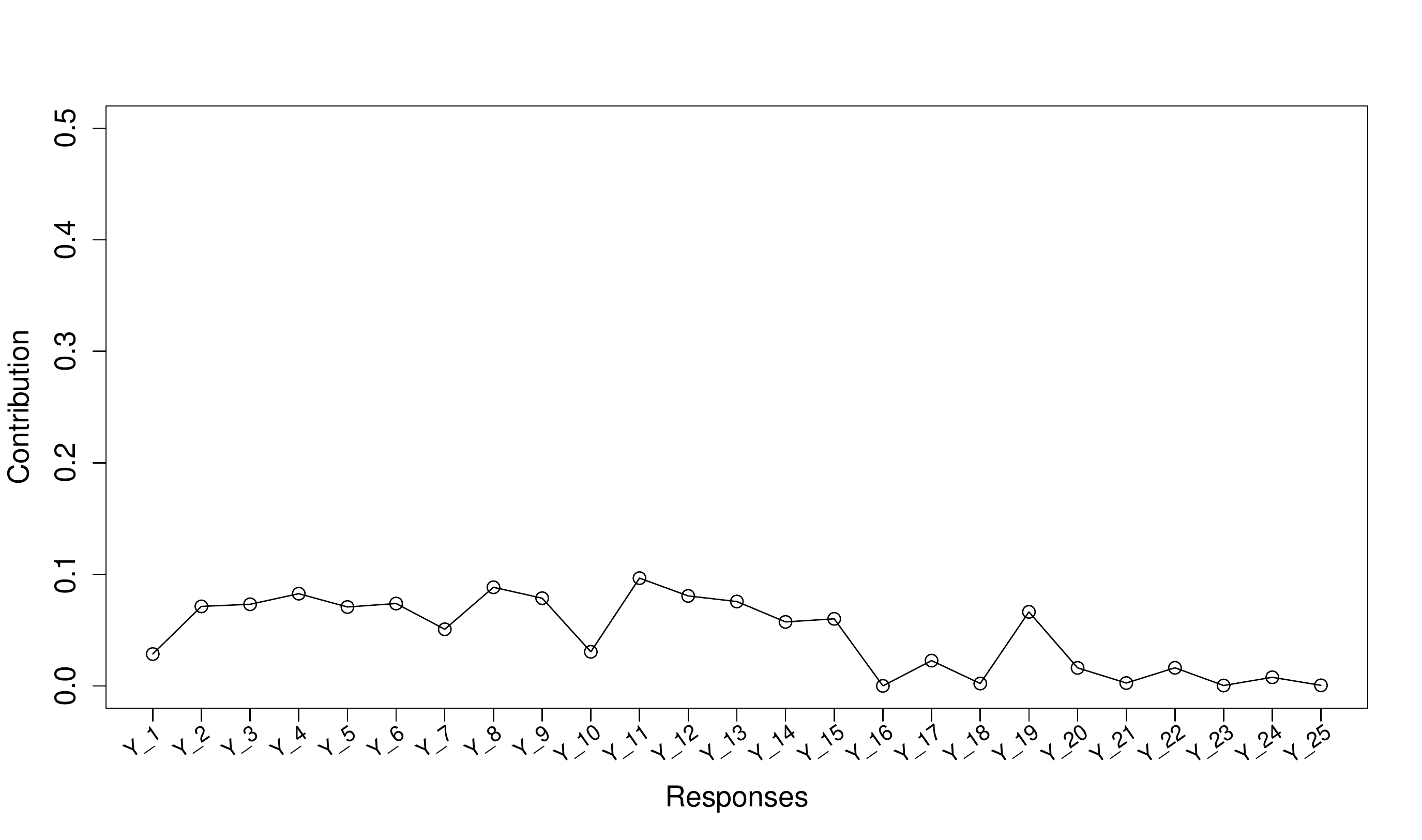}

  \vspace{10pt}
  
  \includegraphics[width=6in]{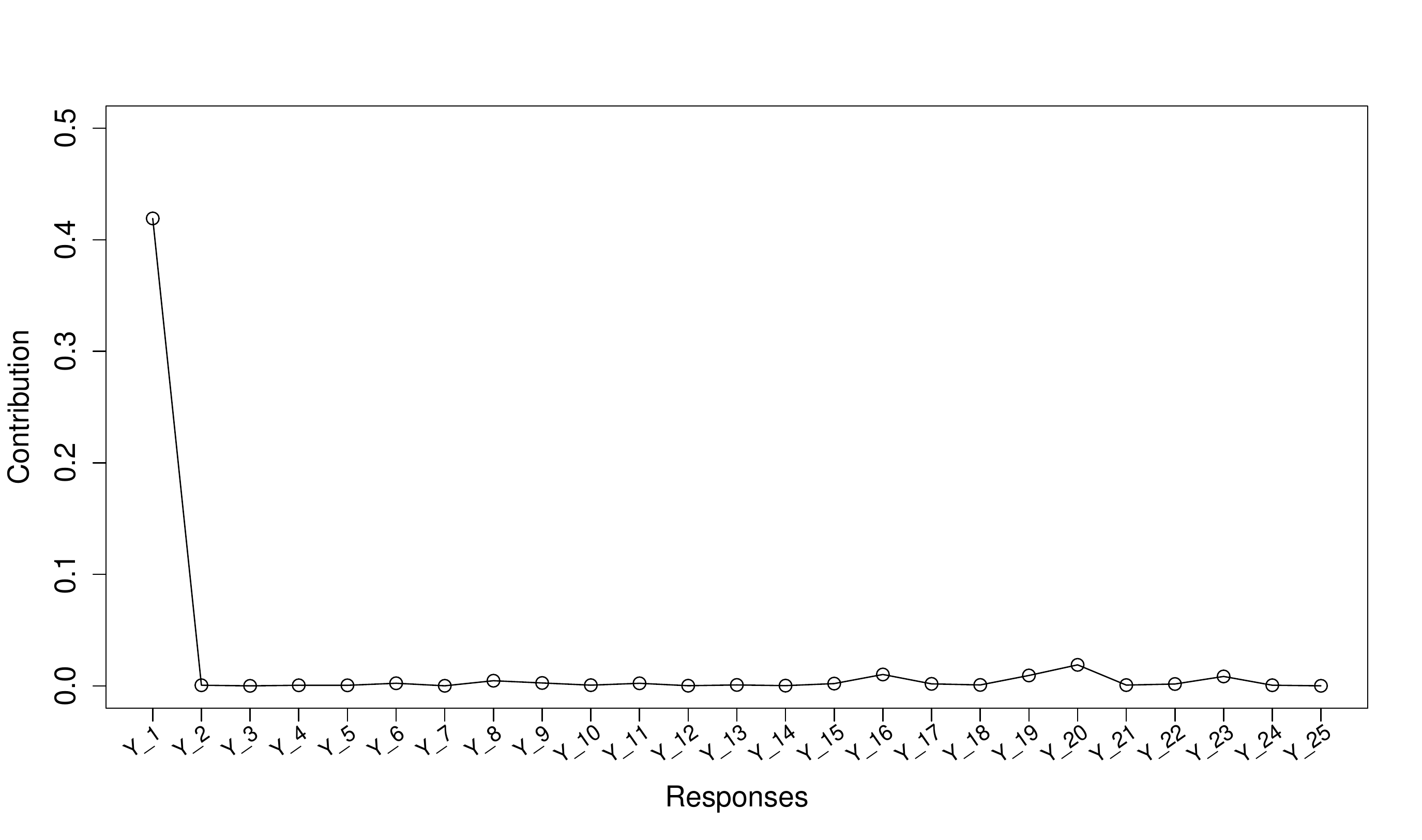}
  \caption
   {Squared correlations between $X^*_{.,100}$ and $Y^*_{.l}$, $l=1,\ldots,25$ (upper) 
and between $X^*_{.,30}$ and $Y^*_{.l}$, $l=1,\ldots,25$ (lower).}
  \label{fig:sim_2_cor_sqr}
\end{figure}

The top panel in Figure \ref{fig:sim_2_cor_sqr} breaks down the 
signal at $X_{100}$ into its squared sample-correlation components, 
$\mathrm{cor}^{2}(X^*_{.100},Y^*_{.l})$. The variable $X_{100}$ appears to be 
modestly associated with the correlated responses $Y_2,\ldots,Y_{15}$,
with the highest pairwise correlation being between $X_{100}$ and $Y_{11}$, even
though the true population correlations between $X_{100}$ and these 
$Y_i$'s are zero. Essentially, we have one modest sample correlation, by chance, 
repeated 14 times due to the correlation between the 14
variables $Y_{2},\ldots,Y_{15}$.
The accumulation of these modest sample correlations
leads to the relatively large contribution for $X_{100}$ in the 
top panel of Figure \ref{fig:sim_2}. The bottom panel of 
Figure \ref{fig:sim_2_cor_sqr} shows the squared sample correlations 
$\mathrm{cor}^{2}(X^*_{.30},Y^*_{.l})$, where 
$\mathrm{cor}^{2}(X^*_{.30},Y^*_{.1})$ reflects a true association. 
As predicted by equation~\eqref{contri_equ}, the indirect associations between
$X_{30}$ and $Y_2,\ldots,Y_{15}$ (due to the modest correlation 
between $Y_1$ and $Y_2,\ldots,Y_{15}$) do not 
play a role in determining the contribution of these response
variables.

\noindent
{\bf Summary of Simulated Data Analyses}
The contribution plot is intended as a \emph{post hoc} investigation of 
an association between multiple explanatory variables and multiple 
response variables, to identify particular explanatory variables that 
may be responsible for the linear association with response variables. 
Our simulated data examples illustrate two main points about the 
contribution plot. First, correlation between explanatory variables 
can widen the peak of a signal, making it difficult to pin-point the 
particular variable(s) driving an association. Second, increasing the 
variance of the contributions, either through correlation between the 
responses or through increasing the number of responses (results not shown),
can obscure 
the signal. However, raising squared correlations to a power can 
counteract this increase in variance and may allow us to identify the 
explanatory variables that are responsible for an association.

\subsection{ADNI Data Results}

The aSPC test of association between the genetic and 
phenotypic variables gives a p-value of 0.0154. The contribution plot 
may therefore be viewed as a {\it post hoc} investigation of the 
significant overall association. To select the power $\alpha_m$ for the 
contribution plot we calculate p-values for SPC($\alpha$) tests. The 
p-values are 0.683, 0.323, 0.062 and 0.008 for $\alpha=1$, 2, 3 and 4, respectively,
leading to $\alpha_m=4$.

\begin{figure}[h]
  \centering
  \includegraphics[width=6in]{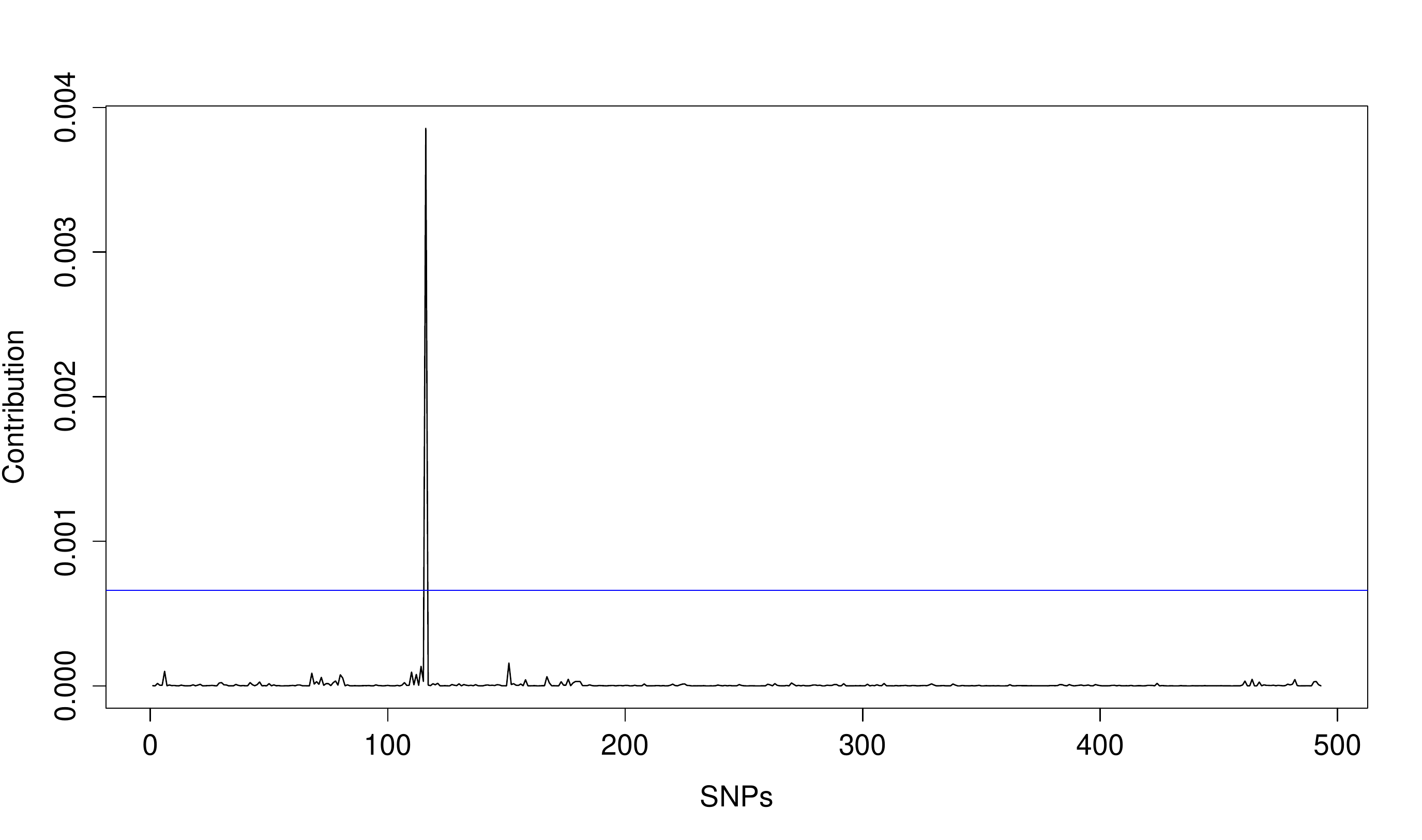}
  \caption
   {Contribution plot of standardized genomic data of 493 SNPs and 56 brain regions with $\alpha=4$.}
  \label{fig:real_data_contribution}
\end{figure}

Figure \ref{fig:real_data_contribution} shows the contribution plot ($\alpha_m=4$). SNPs on the x-axis are sorted by chromosome number and base-pair location. The spike above the permutation-based threshold is a strong signal of a linear association that comes from the SNP \emph{rs16871157} within the \emph{NEDD9} gene on chromosome 6. 

\begin{figure}[ht]
  \centering
  \includegraphics[width=6in]{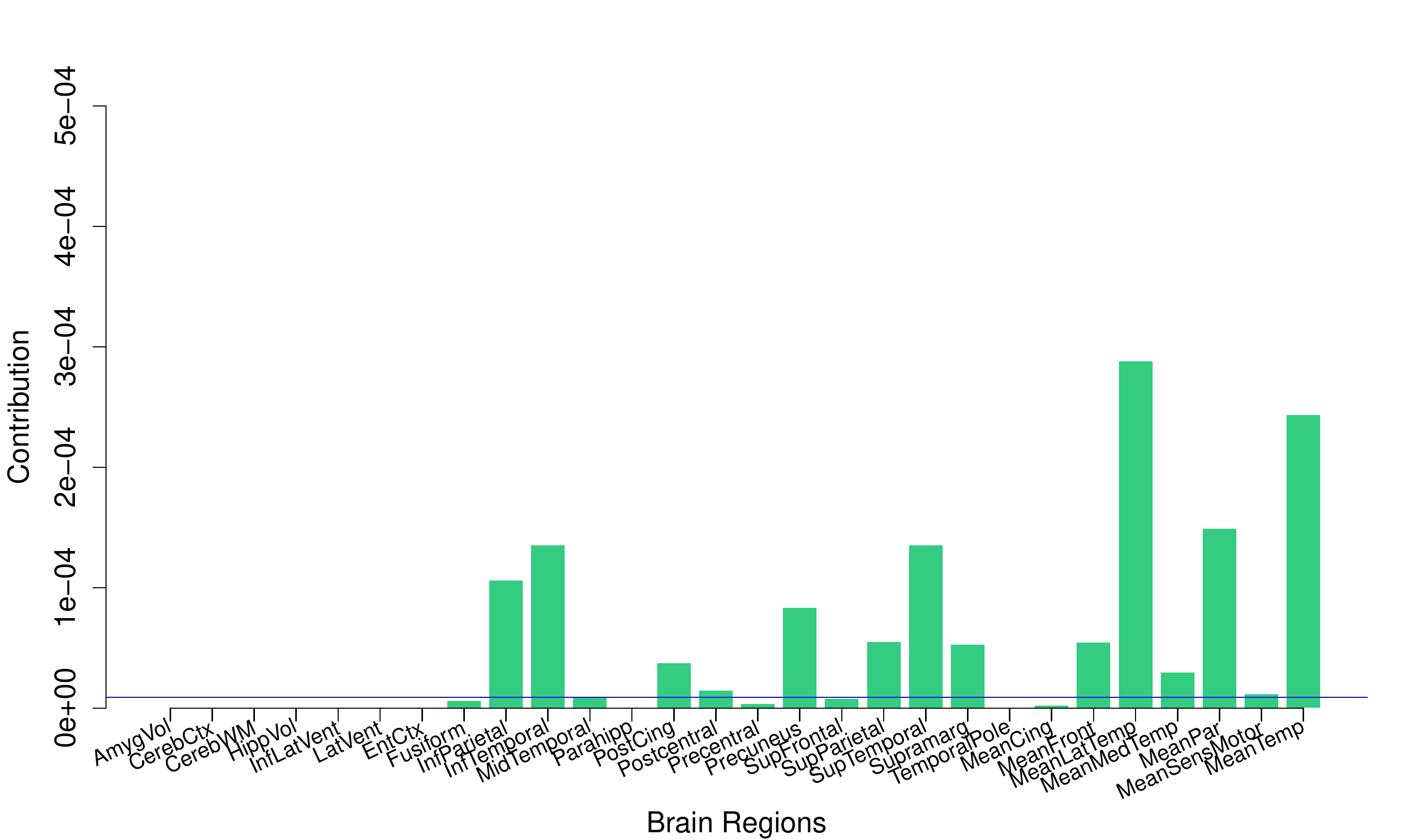}
  
  \vspace{20pt}
  
  \includegraphics[width=6in]{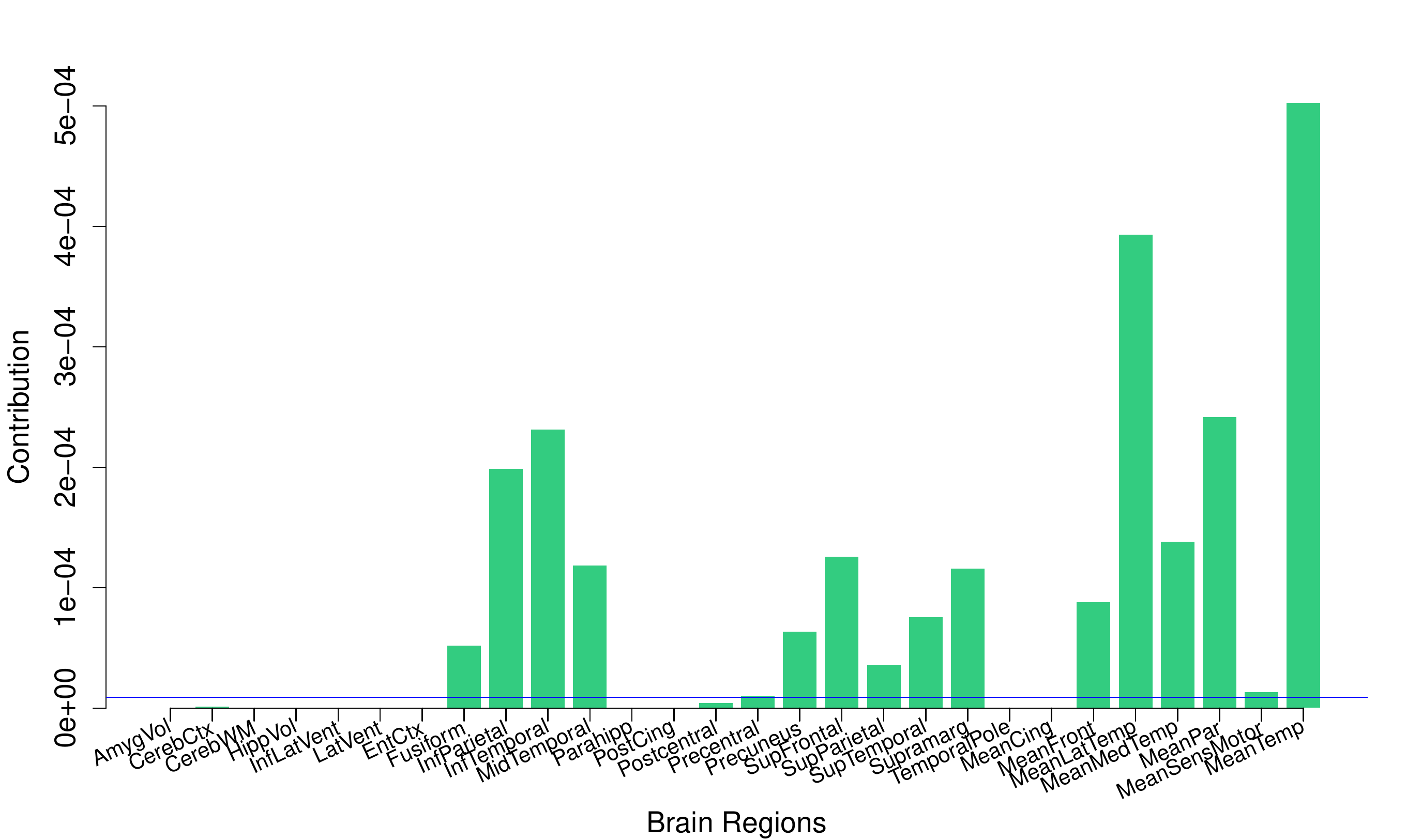}
  \caption
   {Contributions of \emph{rs16871157} to brain regions in the left hemisphere (upper) and the right hemisphere (lower).}
  \label{fig:real_left_right_hemisphere}
\end{figure}

We can further decompose the contribution of \emph{rs16871157} by brain region. 
The results are shown in Figure \ref{fig:real_left_right_hemisphere}, where the 
y-axis represents the individual sample correlation to the power of 8 between 
\emph{rs16871157} and the 56 brain regions. Comparing the two panels of the 
figure, we see that the correlations in the right hemisphere 
are stronger than those in the left hemisphere, but that the patterns of associations 
are very similar. Overall, it appears that \emph{rs16871157} is associated with 
measures of cortical thickness, particularly in the temporal lobe of the brain 
(phenotype MeanTemp).

\begin{figure}[ht]
  \centering
  \includegraphics[width=3.0in]{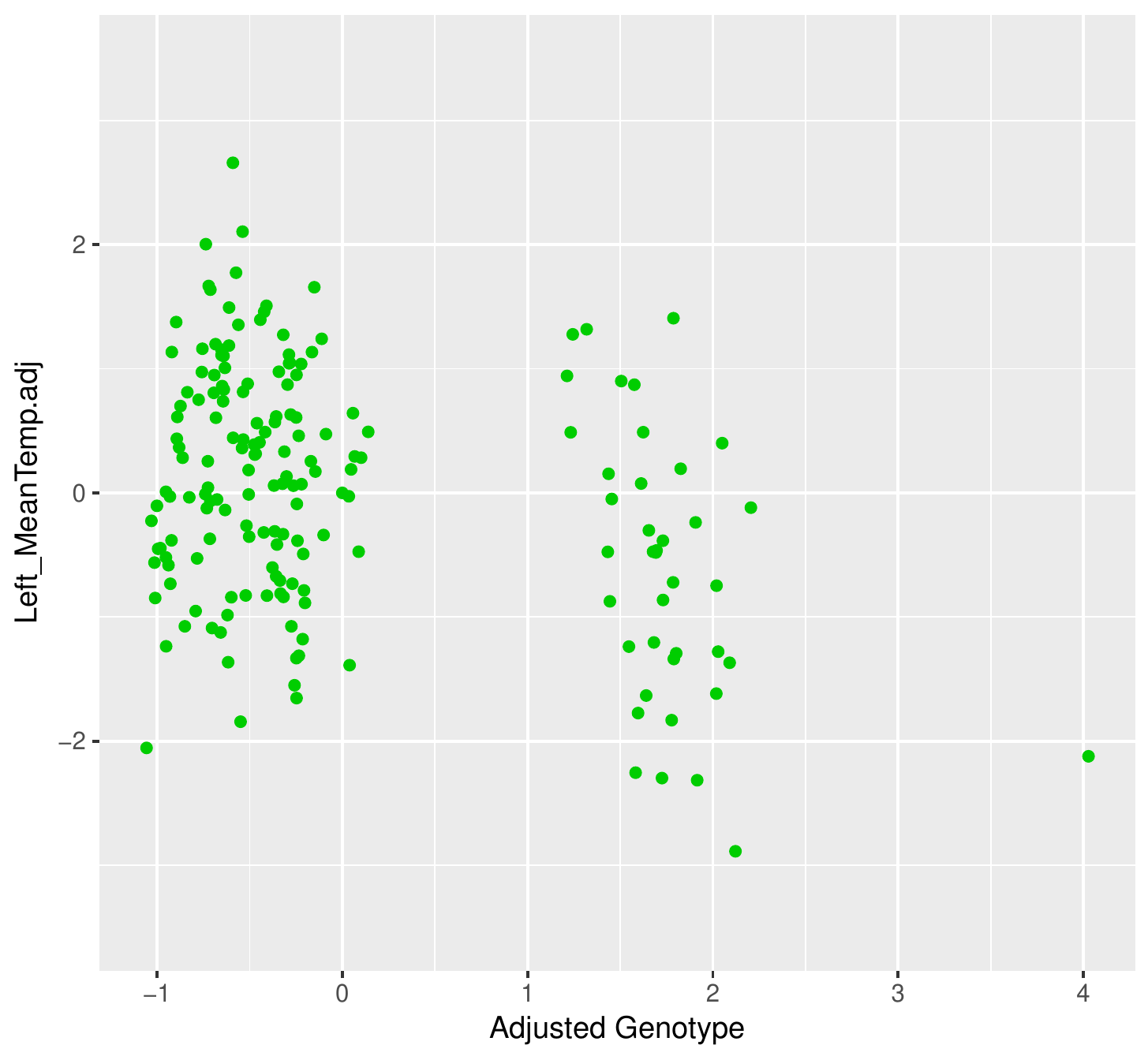}
  \includegraphics[width=3.0in]{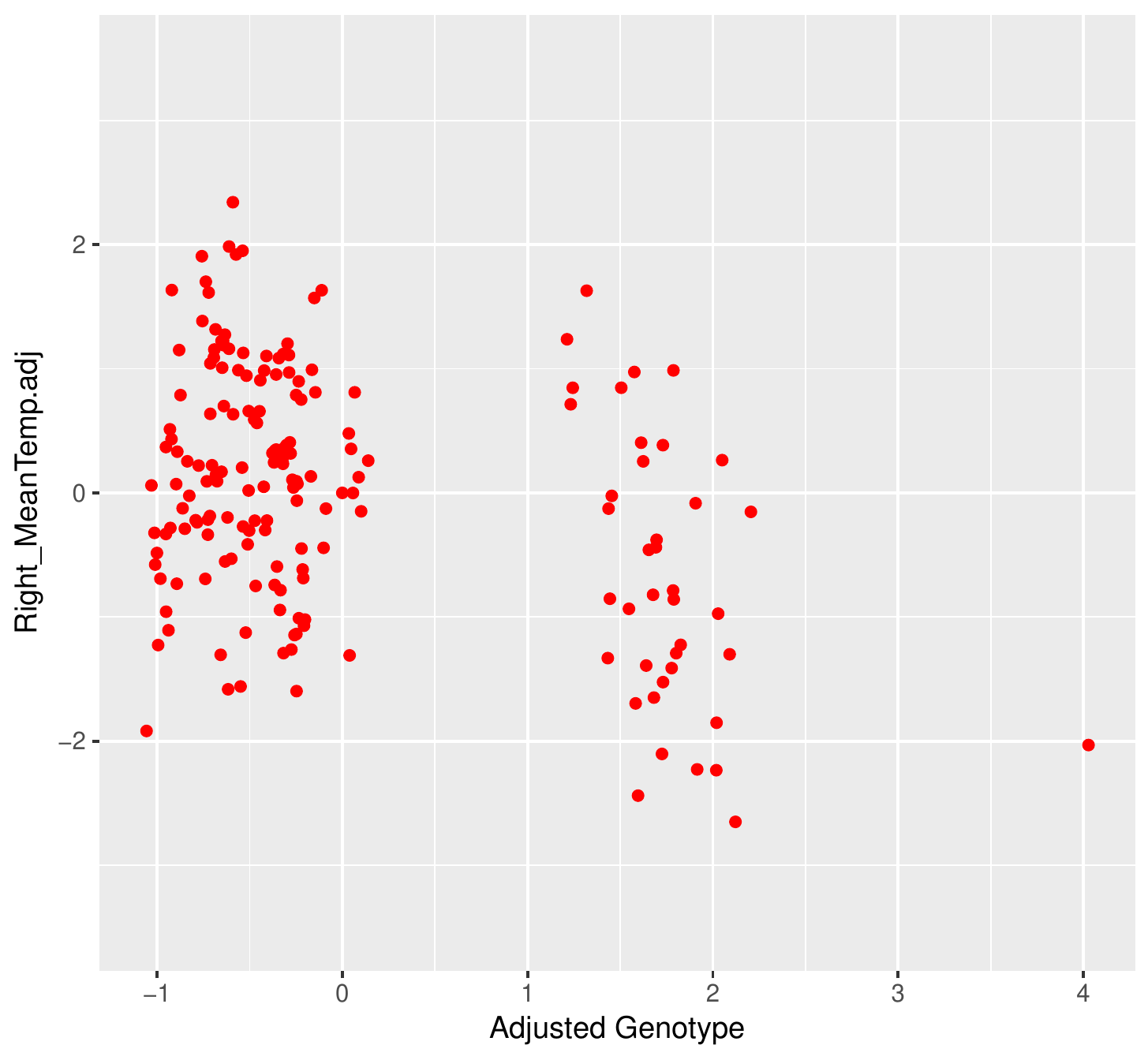}
  
  \vspace{15pt}
  
  \includegraphics[width=3.0in]{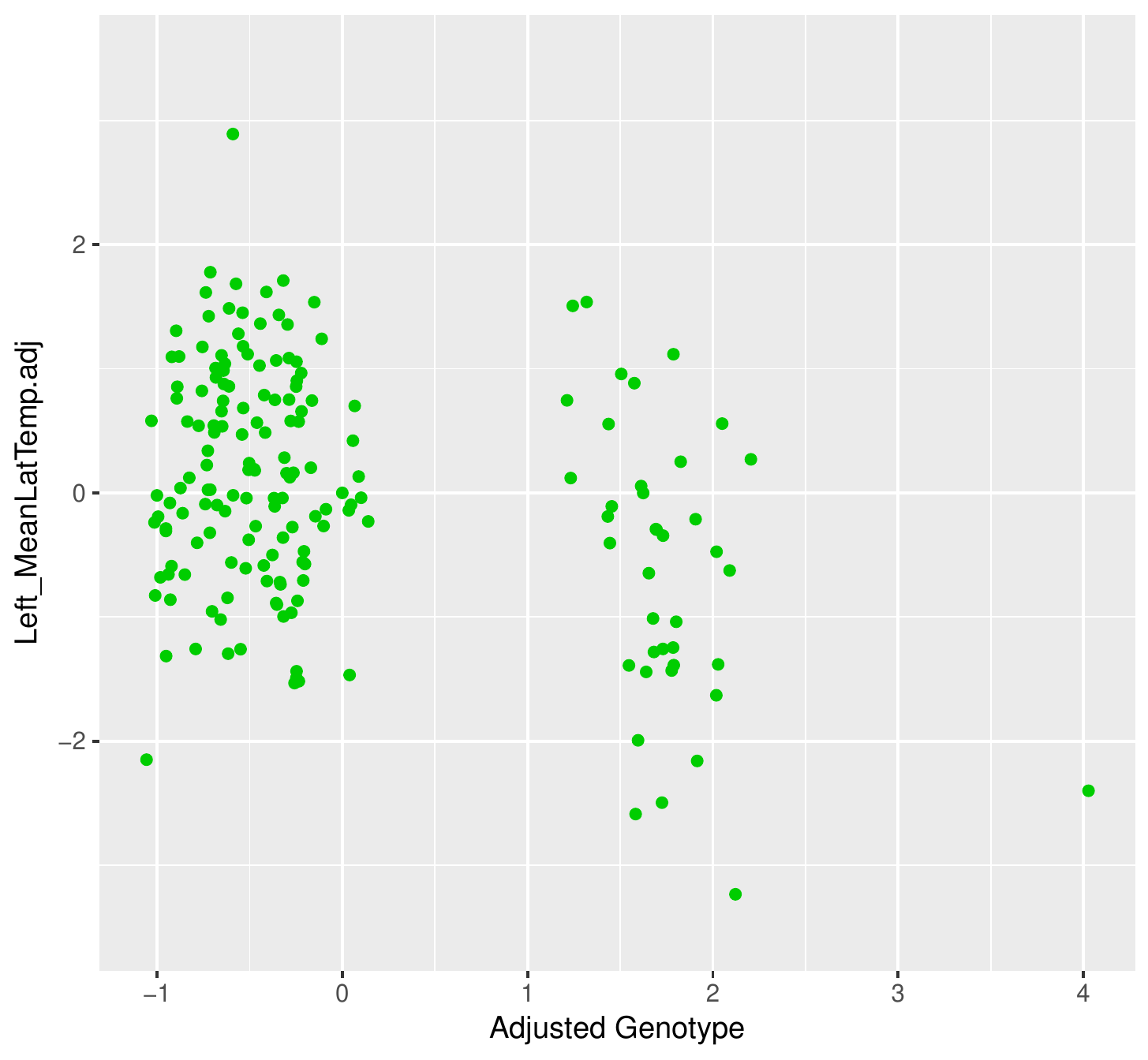}
  \includegraphics[width=3.0in]{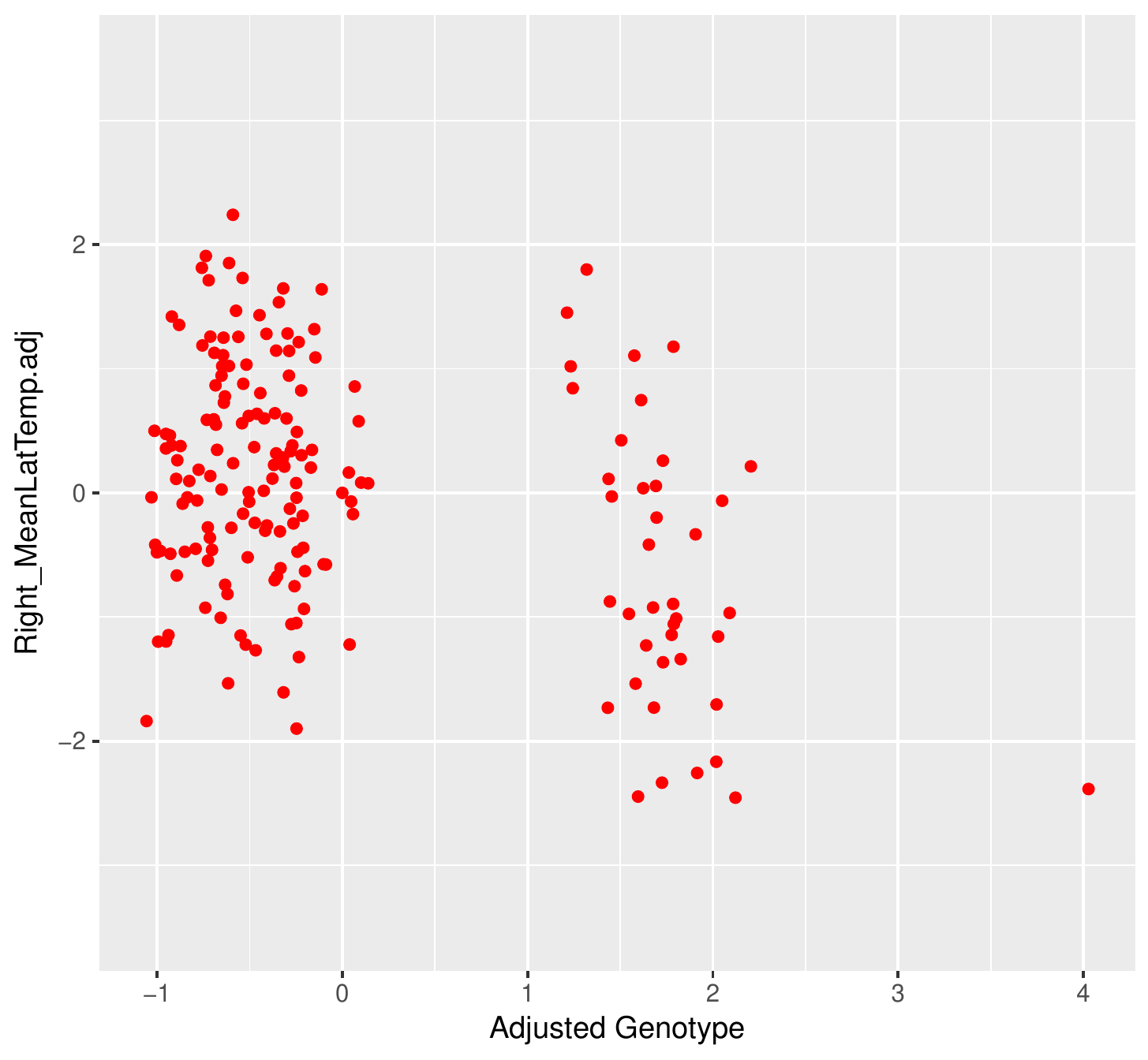}
  \caption
   {Upper panels: Scatterplots of adjusted MeanTemp 
\emph{versus} adjusted \emph{rs16871157} genotype. Lower panels: Scatterplots of MeanLatTemp 
\emph{versus} adjusted \emph{rs16871157} genotype. Adjustments are for ancestry and APOE genotype. The left and right panels are for the left and right hemispheres, respectively.}
  \label{fig:scatter_MeanTemp_plot}
\end{figure}

Scatterplots of adjusted MeanTemp and MeanLatTemp thickness by \emph{rs16871157} 
genotypes are shown in Figure \ref{fig:scatter_MeanTemp_plot} for both the 
left and right hemisphere. In both hemispheres, the distribution of adjusted 
cortical thickness in CN subjects with the variant allele at \emph{rs16871157} 
is shifted towards negative values compared to the distribution for CN subjects 
with two copies of the wild type allele, which is centred at zero. Thus, the 
presence of the variant allele at \emph{rs16871157} is associated with reduced 
cortical thickness in CN subjects.  

\section{Discussion}

Measures of multivariate correlation are used in fields such as 
neurogenetics to find an association between a multivariate phenotype 
and a vector of explanatory variables. After an association is found, 
it may be of interest to identify the explanatory variables that are 
primarily responsible for the signal. In this report we have developed 
such a \emph{post hoc} procedure and applied it to data from the 
ADNI-1 study. The contribution plot decomposes the RV coefficient into 
contributions from each explanatory variable and displays them graphically. 
A significance threshold determined by a permutation procedure may be
added to the plot. Explanatory variables with contributions
above the threshold are considered noteworthy.

Analyses of simulated datasets demonstrated two main points about the 
contribution plot.
\hl{First, localization of the particular variables driving an association
is more difficult when there is correlation between explanatory variables
than when explanatory variables are uncorrelated.
Second, shrinking contributions by raising the component squared-correlations
to a power reduces their variance and can reveal truly-associated
explanatory variables that would otherwise be hidden.}

We applied the contribution plot to the data on CN subjects from ADNI-1. 
The aSPC test for correlation between SNP genotypes and phenotypes of 
brain regions of interest was significant (p=0.0154). The contribution 
plot suggested a sparse signal, driven by a single SNP, \emph{rs16871157}, 
within the \emph{NEDD9} (Neural Precursor Cell Expressed, 
Developmentally Down-Regulated 9) gene on chromosome 6.
\emph{rs16871157} is in an intron of the \emph{NEDD9} gene and has no 
known function. Our results suggest that the variant 
allele at \emph{rs16871157} is associated with reduced cortical 
thickness in CN subjects. Reduced cortical thickness is associated
with symptom severity in mild cognitive impairment and early 
AD patients, and has been observed in CN patients with amyloid 
binding \citep{dickerson2008cortical}.

Much of the research to date on \emph{NEDD9} 
has focussed on the association between 
variation in the gene and different cancers 
\cite[e.g.,][]{izumchenko2009nedd9}, but
the protein product of \emph{NEDD9} is also involved in brain 
development. For example, \cite{vogel2009transforming} found 
that the \emph{NEDD9} protein plays a role in neuronal 
differentiation. In AD research, the SNP \emph{rs760678} in 
\emph{NEDD9} was found to be associated with late-onset AD 
\citep{wang2012nedd9}. However, we note that the phenotypes 
associated with \emph{rs760678} and \emph{rs16871157} are 
quite different (late-onset AD \emph{versus} baseline 
cortical thickness) and the two SNPs are in linkage equilibrium 
in Caucasian populations \cite[estimated $R^2 < 0.01$ in Caucasian 
populations according to the online tool LDlink;][]{mitchell2015LDlink}.

\hl{A reviewer asked about the connection between the contribution plot
and sparse canonical correlation analysis}
\citep[SCCA;][]{parkhomenko2009sparse,witten2009penalized}.
\hl{In canonical correlation analysis} \citep[CCA;][]{Hotelling1936}, 
\hl{the first $k$
pairs of $X$- and $Y$-canonical variates are given by
$\textbf{X}^*M$ and $\textbf{Y}^*L$, where $\textbf{X}^*$ and 
$\textbf{Y}^*$ are column-standardized versions 
of $\textbf{X}$ and $\textbf{Y}$, 
$M$ is the $p\times k$ loading matrix for $\textbf{X}^*$, and $L$ is the $q\times k$ 
loading matrix for $\textbf{Y}^*$. The matrices $M$ and $L$ are obtained by maximizing the 
RV coefficient $RV(\textbf{X}^*M,\textbf{Y}^*L)$} \citep{JosseHolmes16}.
\hl{Our work on the contribution plot suggests an alternative criterion function to maximize, namely
$RV(\textbf{X}^*M,\textbf{Y}^*L|\alpha_m)$, where $\alpha_m$
is the power that minimizes the p-value of the test based
on the generalized RV coefficient $RV(\textbf{X}^*,\textbf{Y}^*|\alpha)$ in 
equation~}(\ref{eqn:modRV}).
\hl{In the resulting canonical pairs, $X$ and $Y$ variables with
low-magnitude correlations are downweighted but not excluded. 
By contrast, in the canonical pairs from SCCA, $X$ and $Y$ 
variables with low-magnitude correlations tend to be excluded.  
Both approaches shrink the loadings in canonical correlation analysis 
but in different ways. Further investigation and comparison of these 
contrasting approaches to shrinking the loadings is a direction for future research. }

The contribution plot can be extended to the case 
where study subjects are differentially weighted. The sample for our study 
was a population sample of CN subjects, and were all equally weighted. If 
instead we had used the entire ADNI-1 sample, which is enriched for MCI and AD 
subjects, we would need to correct for the sampling bias by computing weighted 
covariances or correlations, where the weights are inversely proportional to 
the probability that each subject is included in the sample 
\citep{horvitz1952generalization}. The contribution plot in terms of weighted 
covariance would be of the same form. See \cite{Choi2018}, Appendix A for details.
\hl{Differential weighting also allows one to combine data from different
studies whose sampling designs may differ. Note also that the contributions
depend only on summary statistics (pair-wise correlations), which makes
meta-analysis of summary statistics from multiple studies possible.
Such a meta-analysis 
approach may be useful for data from consortia, such as the ENIGMA Consortium} \cite{ENIGMA}.
Investigating 
the properties of the contribution plot for unequally weighted subjects
\hl{and meta-analysis} is an area for future work.

%
%


%


\section{Statements}

\subsection{Acknowledgment}

The authors would like to thank Elena Szefer for preparing the genetic data.

\subsection{Statement of Ethics}

The authors have no ethical conflicts to disclose.

\subsection{Disclosure Statement}

The authors have no conflicts of interest to declare.

\subsection{Funding Sources}
Data collection and sharing for this project was funded by the Alzheimer's Disease Neuroimaging Initiative
(ADNI) (National Institutes of Health Grant U01 AG024904) and DOD ADNI (Department of Defense award
number W81XWH-12-2-0012). ADNI is funded by the National Institute on Aging, the National Institute of
Biomedical Imaging and Bioengineering, and through generous contributions from the following: AbbVie,
Alzheimer’s Association; Alzheimer’s Drug Discovery Foundation; Araclon Biotech; BioClinica, Inc.; Biogen;
Bristol-Myers Squibb Company; CereSpir, Inc.; Cogstate; Eisai Inc.; Elan Pharmaceuticals, Inc.; Eli Lilly and
Company; EuroImmun; F. Hoffmann-La Roche Ltd and its affiliated company Genentech, Inc.; Fujirebio; GE
Healthcare; IXICO Ltd.; Janssen Alzheimer Immunotherapy Research \& Development, LLC.; Johnson \&
Johnson Pharmaceutical Research \& Development LLC.; Lumosity; Lundbeck; Merck \& Co., Inc.; Meso
Scale Diagnostics, LLC.; NeuroRx Research; Neurotrack Technologies; Novartis Pharmaceuticals
Corporation; Pfizer Inc.; Piramal Imaging; Servier; Takeda Pharmaceutical Company; and Transition
Therapeutics. The Canadian Institutes of Health Research is providing funds to support ADNI clinical sites
in Canada. Private sector contributions are facilitated by the Foundation for the National Institutes of Health
(www.fnih.org). The grantee organization is the Northern California Institute for Research and Education,
and the study is coordinated by the Alzheimer’s Therapeutic Research Institute at the University of Southern
California. ADNI data are disseminated by the Laboratory for Neuro Imaging at the University of Southern
California.

This work is based on JinCheol Choi's MSc thesis supervised by B McNeney and was supported in part by
the Natural Sciences and Engineering Research Council of Canada. 

\subsection{Author Contributions}

JCC developed and implemented the statistical methods, and drafted the manuscript.
DL prepared the phenotype data.
MFB supervised data acquisition and preparation of the phenotype data.
JG developed the statistical methods, supervised preparation of the genotype data and
drafted the manuscript.
BM developed the statistical methods and drafted the manuscript.
All authors revised the manuscript and approved the final version.



\bibliographystyle{frontiersinHLTH&FPHY} 
\bibliography{references}


\section*{Figure captions}


\section*{Appendix: R code to implement the contribution plot}

The contribution plot is intended as a {\it post hoc} procedure that is 
applied after a significant adaptive sum of powered correlations (aSPC) test
\cite{xu2017adaptive}.
The aSPC test is implemented in the {\tt aSPC()} function from the package of the same name
\cite{aSPC17}. {\tt aSPC()} takes two multivariate
data frames and a grid of $\alpha$ values as input, and returns for each $\alpha$ the 
$p$-value for the hypothesis test based on the $RV(X^*,Y^*| \alpha)$ 
statistic (equation~\ref{eqn:modRV}).
For the contribution plot, we choose the power with minimum $p$-value over the grid.

For the chosen $\alpha$ we calculate the contributions of
each explanatory variable to the $RV(X^*,Y^*| \alpha)$ statistic and 
the significance threshold, respectively, with 
the {\tt EstContribution()} and {\tt Threshold()} functions given below.
The contribution plot can then be created with the generic {\tt plot()} 
function in R, as illustrated below.

\begin{verbatim}

EstContribution=function(X, Y, alpha=1){
  # Input: Data matrices X and Y, and the power alpha
  # Output: A vector of contribution of each explanatory variable
  #         to the RV(X,Y| \alpha) statistic
  #
  # 1. Generate a matrix of powered covariances between columns of X and Y
  Cov=(cov(X,Y)^(2))^alpha
  # 2. For each explanatory variable, sum the powered correlations.
  Contr=apply(Cov, 1, sum)
  return(Contr)
}

Threshold=function(X, Y, alpha=1, level=0.95, nrep=100){
  # Input: Data matrices X and Y, the power alpha and the number of 
  #        permutation replicates for the permutation test.
  # Output: The threshold.
  #
  # Initialize a vector to hold max contribution for each permutation 
  maxs = rep(NA,nrep)
  for(i in 1:nrep){
    # record all the maximum contributions based under the estimated 
    # permutation distribution
    maxs[i]=max(EstContribution(X[sample(1:nrow(X)),], Y, alpha=alpha))
    # print the process at every 25%
    if(i%%(0.25*nrep)==0){
      print(paste0("<Obtaining Threshold> ", i/nrep*100, "% done"))
    }
  }
  # obtain the threshold value
  return(quantile(maxs, level))
}

# Call to plot() to plot the contributions and threshold.
plot(EstContribution(X, Y, alpha=1), type='l', main="X and Y (alpha=1)", 
     xlab="Explanatory Variables", ylab="Contribution",
     cex=1.3, cex.lab=1.3, cex.axis=1.3, cex.main=2.5, cex.sub=1.3)
abline(a=Threshold(X, Y, alpha=1),b=0,col="blue")
\end{verbatim}

\end{document}